----------
X-Sun-Data-Type: default
X-Sun-Data-Description: default
X-Sun-Data-Name: draft4.tex
X-Sun-Content-Lines: 420

\documentstyle[preprint,aps]{revtex}
\begin{document}
\draft

\title{Conductance of an Artificial Atom in Strong Magnetic Fields}
\author{O. Klein, C. de C. Chamon, D. Tang, D.M.~Abusch-Magder, X.-G.~Wen, and
M.A.~Kastner}
\address{Physics Department, Massachusetts Institute of Technology,\\ 77
Massachusetts Ave., Cambridge MA 02139}
\author{S.J. Wind}
\address{IBM T.J. Watson Research Center, Yorktown Heights, New York 10598}
\date{Submitted to Phys. Rev. Lett. 6 July 1994}
\maketitle
\begin{abstract}

The conductance resulting from resonant tunneling through a droplet of
$N \sim 30$ electrons is used to measure its chemical potential
$\mu_N$. Abrupt shifts of $\mu_N$ occur at sharply defined values of
the magnetic field, at which the state of the droplet changes. These
are used to study part of the phase-diagram of the droplet in strong
magnetic fields; we find evidence for a new phase in the spin
polarized regime. We make a detailed comparison between theory and
experiment: Hartree-Fock provides a quantitative description of the
measurements when both spin-split states of the lowest orbital Landau
level are occupied and a qualitative one in the spin polarized regime.

\end{abstract}

\narrowtext

Recent experiments \cite{mceuen:prl,ashoori:prl} have demonstrated the
possibility of measuring the chemical potential $\mu_N$ of a droplet
of $N$ electrons confined by an external potential, an artificial
atom. Abrupt shifts of $\mu_N$ occur at values of the magnetic field
$B$ at which the ground state (GS) of the droplet changes. These
results have stimulated calculations of the $B$-$N$ phase diagram, in
which each phase is designated by the quantum numbers of the GS; the
changes in $\mu_N(B)$ happen at the phase boundaries. Because exact
numerical calculations are possible only for $N\leq 6$
\cite{yang,pfannkuche}, approximate methods
\cite{mceuen:pb,mcdonald,chamon} have been used for larger $N$ to
account for electron-electron interactions. The strong magnetic field
regime is the appealing place to test these approximations, because
the most intriguing aspects of the phase-diagram occur at these
fields. In particular, MacDonald {\it et al.} \cite{mcdonald}, and
Chamon {\it et al.} \cite{chamon} have independently predicted the
existence of new phases of a spin polarized droplet in a parabolic
potential. These phases are especially interesting because any
transition in the spin polarized regime is the consequence of
many-body phenomena that cannot be explained by a single-electron
picture.

In this Letter, we present detailed measurements of a portion of the
phase diagram in strong magnetic field. We propose a new systematic
approach for comparing the experimental results with the models. Using
this approach, we find that Hartree-Fock (HF) \cite{mcdonald,chamon}
provides a quantitative description when both spin states of the
lowest orbital Landau level (LL) are occupied, whereas a
semi-classical model (SC) \cite{mceuen:pb} does not, indicating that
exchange plays an important role. We also find evidence for a new
phase in the spin polarized regime, which is described qualitatively
by HF.

The device that we study is of the type described by Meirav {\it et
al.} \cite{meirav}. It consists of a two-dimensional electron gas
(2DEG) in an inverted GaAs/Al$_{x}$Ga$_{1-x}$As heterostructure with
electrostatic gates above and below it. The bottom gate is a highly
conducting substrate of $n^+$ doped GaAs. A positive bias, $V_g$,
applied to the bottom gate varies the density of the 2DEG. On the top
surface of undoped GaAs, two metallic (TiAu) gates are
lithographically patterned with a double constriction. Applying a
negative bias to these top gates depletes the 2DEG 100nm underneath
them, confining the electrons to an island between the constrictions.
Current flows through the resulting electron droplet via the tunnel
barriers caused by the constrictions. The top gate geometry of the
device under investigation has been examined with an Atomic Force
Microscope. The constrictions are poorly defined, but we estimate that
the region between them is roughly $500\times500$nm$^2$.  According to
the simulation of the device by Kumar {\it et al.} \cite{kumar}, the
external confinement potential of the droplet is approximately
parabolic. Although all results presented here are for this one
structure, we have observed similar features in samples of different
geometries.

The negative bias on the top gate is maintained constant during the
experiment and the bottom gate voltage is varied in a narrow range
near $V_g=160\pm 1$mV, for which the electron density of the 2DEG
regions outside the constriction is almost constant at $1.3 \pm
0.01\times 10^{11}$cm$^{-2}$. The conductance $G$ of the device as a
function of $V_g$ at $B=0$T is shown in the lower inset of Fig.1. It
consists of quasi-periodic sharp peaks ($\Delta V_g =1.2$mV), crudely
described by the coulomb blockade mechanism \cite{beenakker}. In this
model, when the bottom gate voltage is set between peaks, transport is
suppressed by the charging energy $U\sim0.66$meV necessary to add an
electron to the droplet. Each period thus corresponds to the addition
of one electron to the droplet.  At resonance, the electrochemical
potential of the droplet, $\mu_N -e \alpha V_g$, is aligned with the
Fermi energy of the leads and current flows; current requires a
fluctuation of the charge on the droplet. Thus, the value of $V_g$ at
which the peak occurs provides a measure of $\mu_N$. At T$=0$,
$\mu_N=E_N-E_{N-1}$, where $E_N$ is the energy of the $N$-electron GS.

We begin by considering the effect of magnetic field on a single
conductance peak. That is, we measure $\mu_N(B)$ at constant $N$. The
value of the gate voltage at which the $N$th conductance peak occurs
is plotted as a function of magnetic field between 1 and 5T in Fig.1.
McEuen {\it et al.} recognized that the change in behavior near 1.6T
results from the depopulation of all but the lowest orbital LL
\cite{mceuen:pb}. The step-like behavior of the peak position above
1.6T can be thought of as resulting from the transfer of electrons
between the two spin-split states of the lowest orbital LL
\cite{mceuen:pb}. Each step corresponds to a change in the quantum
numbers of the GS, for example, the total spin of the droplet.  The
number of steps above 1.6T is proportional to $N$, the number of
electrons in the droplet, but the proportionality constant depends on
the shape of the charge distribution.

A novel way to characterize the data in Fig.1 is to examine the
separation in $B$ of the upward steps. The peak conductance as a
function of $B$ has a sharp minimum at each of these steps
\cite{mceuen:prl}. The $n$th minimum precisely determines $B_n$, the
field for the $n$th step. (The $B_n$ are indicated by arrows in
Fig.1.) We plot in Fig.2a the quantity $(B_n-B_{n-1})^{-1}$ as a
function of $B_n$. Because each step corresponds to the flip of a
spin, one may think of $(B_n-B_{n-1})^{-1}$ as being roughly
proportional to the spin susceptibility. A fit to the form $y(B)=y_o
[(B-B^{\prime}) / B^{\prime}]^{\epsilon}$ gives $B^{\prime} = 1.7 \pm
0.02$T and $\epsilon=-0.41 \pm 0.06$ for our data; the solid curve in
Fig.2a shows the fit. The same functional form also fits the
experimental data for two other devices with different geometries
($500\times700$nm$^2$ and $450\times900$nm$^2$) and larger numbers of
steps ($\sim 25$ and $35$); we find $\epsilon=-0.37\pm0.1$ for all
three devices.

Plotted in Fig.2c is the result obtained when $(B_n-B_{n-1})^{-1}$ is
determined using $\mu_N$ of the SC model of McEuen {\it et
al.}\cite{mceuen:pb}. For a parabolic potential with cylindrical
symmetry, $V(r) = m^* \omega_0^2 r^2/2$, the SC spatial density of
electrons $\rho(r)$ is approximately that of classical electrons in
zero magnetic field $\rho(0) \sqrt{1-(r/R)^2}$, except near $r$
corresponding to integer filling factors $\nu=2\pi\ell^2\rho(r)$ where
the electrons form an incompressible liquid. ($m^*$ is the effective
mass of electrons in GaAs, $\omega_0$ is the oscillator frequency, $R$
is the radius of the droplet, and $\ell$ is the magnetic length.) In
this picture, the change in behavior at $B=1.6$T corresponds to a
filling factor $\nu=2$ at $r=0$. At fixed $B$, $\rho(r)$ is uniquely
determined by $N$ and $\hbar \omega_0$. We adjust $N$ so that the
calculated $\mu_N(B)$ has the same number of steps as observed
experimentally. With $N$ fixed, we adjust $\hbar \omega_0$ to match
the value of $B^{\prime}=1.7$T, at which the transfer of electrons
between the two spin states of the lowest LL begins in our experiment.
Using this procedure, we find $N=42$ and $\hbar \omega_0 =1.8$meV. As
seen in Fig.2c the SC model predicts values of $(B_n-B_{n-1})^{-1}$
which are roughly the same size as the measured ones.  However, it
does not predict the upward curvature of $(B_n-B_{n-1})^{-1}$ near
$B^{\prime}$. In the SC calculation, the last spin flip occurs at
$B_I=3.23$T$=1.9 B^{\prime}$. In fact, the SC value of the ratio
$B_I/B^{\prime}$ is almost independent of $N$ and $\hbar\omega_0$.
Although we find a step at $2 B^{\prime}$ in our measurement, there is
also an additional step at $B_c = 3.75$T$ = 2.2 B^{\prime}$ (filled
circle in Fig.1) not predicted by the SC model.  For all devices
studied we find a step at $2 B^\prime$, marking the complete
depopulation of the higher energy spin state, and a step at larger
field, in the spin polarized regime \cite{nbex}. For one device, we
have explored the phase diagram beyond $2.7 B^\prime$ and have found
evidence for other steps \cite{olivier}.

The step at $B_c$ behaves in a way that is very different from those
between $B^\prime$ and $2 B^\prime$. By examining successive peaks in
$G$ vs. $V_g$ (lower inset of Fig.1), i.e. probing the droplet at
successive $N$, we find that each step in $\mu_N$ (Fig.1) shifts to
higher $B$ when another electron is added to the droplet. We have
averaged the shift over four consecutive conductance peaks and have
plotted its inverse $[B_n(N) - B_n(N-1)]^{-1}$ in Fig.3a for each of
the steps in Fig.1. This quantity measures the slope of the phase
boundary ($\partial N/\partial B$).  It is clear from Fig.3a, that the
step at $B_c$ has a larger slope than those at lower $B$.

The temperature dependence of the step at $B_c$ is also peculiar.
Figs.4a and b show that the features between 1.7 and 3.4T disappear by
500mK as $T$ is increased. This behavior is now well understood
\cite{mceuen:prl}. In clear contrast, the height of the step at
$B_c=3.75$T does not change with temperature up to 800mK, our
measurement limit.

To compare our results with a more sophisticated theory, we have
performed a HF calculation of $\mu_N(B)$, choosing the states of the
symmetric gauge as the complete basis set, with the Hilbert space
truncated to the two spin states of the lowest LL. Because of
exchange, the HF $\rho(r)$ is more compact than the SC $\rho(r)$ with
larger incompressible regions, smaller compressible regions and a more
rapid decrease with $r$ near the edge of the droplet \cite{chamon}.

In Fig.2b, we plot $(B_n-B_{n-1})^{-1}$ as a function $B_n$ extracted
from $\mu_N(B)$ for the HF model with $N=27$ electrons and $\hbar
\omega_0=2.1$meV. As with the SC model, these parameters are chosen to
match the number of steps in the peak position and the experimental
value of $B^{\prime}$, respectively. Like the value of $\hbar
\omega_0$ used for the SC model ($1.8$meV), the $\hbar \omega_0$ that
fits the HF calculation is in agreement with the value $2\pm1$meV
calculated by Kumar {\it et al.} from the sample geometry
\cite{kumar}. The HF value of $N=27$, however, is different from that
($N=42$) which fits the SC model. This discrepancy is a result of the
difference in shape of the charge distribution. In the HF model, the
number of steps in $\mu_N(B)$ is equal to $N/2$ because the two spin
states are equally occupied at $B^{\prime}$, and half the electrons
flip their spin as the field is increased between $B^\prime$ and
$B_I$.

It is obvious from Fig.2b that the HF calculation is in excellent
quantitative agreement with the experiment (Fig.2a). This is
particularly impressive since there are no other fitting parameters
once $N$ and $\hbar \omega_0$ are fixed. In particular, the HF model
predicts correctly the apparent divergence of $(B_n-B_{n-1})^{-1}$
near $B^{\prime}$, in clear contrast with the SC model. A fit to the
HF results with $y(B)$ gives $\epsilon=-0.43\pm 0.03$, which is the
same as the experimental value within the errors. The apparent
divergence of $(B_n-B_{n-1})^{-1}$ in Fig.2a suggests that because of
exchange the two spin states of the lowest LL are equally occupied at
$B^{\prime}$ in our droplet. This is consistent with another
experimental observation: a new step in $\mu_N(B)$ is added between
$B^\prime$ and $2 B^\prime$ for every two electrons added, implying
that the two spin states of the lowest LL of our droplet are equally
populated with increasing $N$.

The HF calculation predicts that the last spin flip occurs at
$B_I=3.15$T$=1.85B^{\prime}$. Like the SC model, the HF ratio
$B_I/B^{\prime}$ is nearly independent of $N$ or $\hbar \omega_0$.
Thus, above $B_I$ the droplet is spin polarized. MacDonald {\it et
al.} \cite{mcdonald}, and Chamon {\it et al.} \cite{chamon} showed
that there exists a region in the $B$-$N$ phase diagram (sketched in
the upper inset of Fig.1) in which, for $N<N_c\sim 100$, the GS of the
spin polarized droplet is the maximum density droplet (MDD). In the
MDD state, all the single-particle eigenstates of angular momentum
index $m= 0,1, ..N-1$ are occupied, leading to an approximately
constant $\rho(r)$ in the droplet.  The MDD is of course the GS of
non-interacting electrons at high $B$, but surprisingly it is also the
GS in a region of $B$-$N$ even in the presence of repulsive
interactions \cite{mcdonald,chamon}.

 With increasing magnetic field the radius of the MDD decreases, the
electrons get closer together, and the interaction energy eventually
favors a larger area droplet. HF \cite{chamon} predicts that, at
$B_c$, the edge undergoes a reconstruction and electrons form an
annulus at a distance $\sim 2 \ell$ away from the central droplet,
causing an abrupt upward shift of $\mu_N$ at $B_c$ of roughly the same
height as the step at $B_I$ \cite{chamon}. In the HF calculation,
$B_c/B_I$ is almost independent of $\hbar\omega_0$, but it decreases
with increasing $N$ \cite{mcdonald,chamon} for $N<N_c$.

The excellent quantitative agreement between HF and the experiment for
$B\leq 2 B^\prime$, strongly suggests that the MDD is formed in our
experiment above $2 B^{\prime}$. The HF calculation predicts that the
transition to the reconstructed droplet occurs at $B_c=4.21$T for our
droplet, a value larger than the one observed experimentally. In this
regard, it is important to bear in mind that although the HF energy
of the MDD is exact because the MDD is an exact eigenstate of the
many-body Hamiltonian \cite{mcdonald}, the HF energy of the
reconstructed droplet is only variational. Therefore, the calculated
value of $B_c$ is an upper bound on the true transition field. Indeed,
an exact calculation for small $N$ \cite{yang,pfannkuche} shows that
the HF model overestimates $B_c$.

Turning to the slopes of the phase boundaries,
one sees in Fig.3 that $[B_n(N)-B_n(N-1)]^{-1}$ from HF ($\sim 2.2
\times 10^{-3}$G$^{-1}$) agrees fairly well with experiment ($3
\pm 1 \times 10^{-3}$G$^{-1}$) between $B^\prime$ and $2 B^\prime$.
However, at $B_c$, the HF value $3.2 \times 10^{-3}$G$^{-1}$, is
smaller than the experimental value, $8 \pm 1.5 \times
10^{-3}$G$^{-1}$. The quantities $[B_n(N)-B_n(N-1)]^{-1}$ at $B_I$ and
$B_c$ are the slopes of the phase boundaries in the $B$-$N$ phase
diagram between which the MDD is the GS. The fact that
$[B_n(N)-B_n(N-1)]^{-1}$ is larger at $B_c$ than at $B_I$ suggests
that the MDD does not exist above some $N_c$ \cite{mcdonald,chamon}.
The experimental observation of both a larger value of
$[B_n(N)-B_n(N-1)]^{-1}$ at $B_c$ and a smaller value of $B_c$ than
the ones predicted by HF suggests that $N_c$ is smaller than
predicted by HF.

We have extended the HF calculation to obtain excited states and thus
study the temperature dependence of $\mu_N$. We find that the HF
excitation spectrum (proportional to the height of the $\mu_N$ steps)
has an energy scale 4 times larger than the experimental one over the
entire magnetic field range. Nonetheless, the steps in the region
between $B^{\prime}$ and $B_I$ are predicted to wash out more rapidly
with increasing $T$ than the one at $B_c$ in agreement with
observation (Fig.4).

Finally, we note that HF also describes the $B$ dependence of the
conductance peak height \cite{olivier}. HF predicts the experimentally
observed \cite{olivier} decrease in peak height just below $B_c$
followed by an increase for $B$ above $B_c$. The increase above $B_c$
is ascribed to the reduced separation between the edge of the droplet
and the leads when the annulus is formed.

The failure of HF to predict the size of the magnetic field window in
which the MDD is the GS (Fig.2b) and the dependence of $B_c$ on
$N$ (Fig.3b) may indicate that correlations are playing an important
role in this transition. The downward step at about 3.5T (Fig.1) is
also reminiscent of features predicted to result from correlations
\cite{yang}.

In conclusion, we have made a detailed study of the conductance peak
positions in strong magnetic fields. We have focused on that part of
the phase diagram in which only the lowest orbital LL with its two
spin-split states are occupied. By looking at the increase in magnetic
field required to flip each successive spin, we are able to make a
quantitative comparison between experiment and theory. We find that HF
is in excellent quantitative agreement with experiment at low field.
However, when the droplet is spin polarized a new transition occurs
which is only qualitatively described by HF.

We are grateful to Udi Meirav who made the samples. We thank R.C.
Ashoori, D.B. Chklovskii, K.A. Matveev and N.S. Wingreen for many
useful discussions. We also thank Nathan and Paul Belk for their help
in the experiment. This work was supported by NSF Grant No. ECS
9203427 and by the U.S. Joint Services Electronics Program under
Contract No. DAALL03-93-C-0001.


\begin{thebibliography}{10}
\bibitem{nbex}
We point out that while the results reported here are completely reversible and
  continuous, we have, from time to time, observed hysteresis and discontinuous
  behavior in strong magnetic fields.
\end{thebibliography}

\begin{figure} \caption {Upper Inset: $B$-$N$ phase diagram of the
droplet. The boundaries corresponding to a change of the total spin of
the droplet in the $2>\nu>1$ regime are omitted; the MDD domain of
stability is limited on one side by $B_I(N)$, the boundary of the spin
polarized phase, and on the other side by $B_c(N)$, where there is a
reconstruction of the charge density.  Above $N_c$ the MDD phase is
terminated. Lower Inset: Conductance through the island as a function
of the bottom gate voltage at $B=0$T. Main: Position of the $N$th
conductance peak as a function of $B$ at $T=100$mK. We have used a
constant factor $\alpha=0.55$ to convert the bottom gate voltage scale
to energy [1,5]. The arrows indicate the minima of the conductance
peak height. $B_n$ is the field for the $n$th minimum above 1.6T, with
$n=\{1,...,14\}$.} \end{figure}

\begin{figure} \caption {(a) $(B_n-B_{n-1})^{-1}$ vs. the $B_n$
obtained from Fig.1.  The error bars represent the spread of the data
when the analysis is repeated for other conductance peaks on the same
device. (b) and (c) results obtained with the HF ($N=27$ and
$\hbar\omega_0=2.1$meV) and SC ($N=42$ and $\hbar\omega_0=1.8$meV)
calculations. The solid line is a fit with $y(B)=y_o [(B-B^{\prime}) /
B^{\prime}]^\epsilon$ where $B^{\prime} = 1.7 \pm 0.02$T and
$\epsilon=-0.41 \pm 0.06$ for the experiment and $\epsilon=-0.43\pm
0.03$ for the HF. $B_I$ indicates the field onset of the spin
polarized regime in both models.  The solid circle indicates $B_c$.
The dashed line in (c) is the constant interaction model [5] (the
scale is expanded by a factor 10).} \end{figure}

\begin{figure} \caption {(a) $[B_n(N)-B_n(N-1)]^{-1}$ is the slope of
the phase boundary; it is measured by looking at the shift in
$B$-field of the same $B_n$ between two adjacent conductance peaks (or
equivalently when an electron is added to the droplet). The plotted
value is the average over 4 consecutive peaks and the error bars are
the standard deviations (b) HF and SC values of
$[B_n(N)-B_n(N-1)]^{-1}$ measured from the simulated $\mu_N(B)$.}
\end{figure}

\begin{figure} \caption {Magnetic field dependence of the peak
position at 100 and 500mK. (a) and (b) show the behavior below $2
B^{\prime}$, while (c) and (d) show it above $2 B^{\prime}$. In each
figure, the peak position is offset for clarity.} \end{figure}

\end{document}